\documentclass[prl,twocolumn,showpacs,superscriptaddress]{revtex4-1}
\usepackage{dcolumn}
\usepackage{bm}
\usepackage{graphicx}
\usepackage{amsmath}
\usepackage{amssymb}

\begin{document}
\title{
Evolution of the Kondo lattice and non-Fermi liquid excitations
in a heavy-fermion metal}

\author{S. Seiro}\altaffiliation[Present address: ]{Institute for
Solid State Physics, IFW-Dresden, Helmholtzstrasse 20,
01069 Dresden, Germany}
\affiliation{Max Planck Institute for Chemical Physics of Solids,
01187 Dresden, Germany}
\author{L. Jiao} \affiliation{Max Planck Institute for Chemical
Physics of Solids, 01187 Dresden, Germany}
\author{S. Kirchner}
\affiliation{Center for Correlated Matter, Zhejiang University,
Hangzhou, China}
\author{S. Hartmann} \affiliation{Fraunhofer Institute for
Photonic Microsystems, 01109 Dresden, Germany}
\author{S. Friedemann} \affiliation{School of Physics,
University of Bristol, Bristol BS8 1TH, UK}
\author{C. Krellner} \affiliation{Institute of Physics,
Goethe-University Frankfurt, 60438 Frankfurt/Main, Germany}
\author{C. Geibel}
\affiliation{Max Planck Institute for Chemical Physics of Solids,
01187 Dresden, Germany}
\author{Q. Si} \affiliation{Department of Physics and Astronomy,
Rice University, Houston, Texas 77005, USA}
\author{F. Steglich}
\affiliation{Max Planck Institute for Chemical Physics of Solids,
01187 Dresden, Germany}
\author{S. Wirth}\thanks{corresponding author}
\affiliation{Max Planck Institute for Chemical Physics of Solids,
01187 Dresden, Germany}
\date{\today}\maketitle

\textbf{
Strong electron correlations can give rise to extraordinary
properties of metals with renormalized quasiparticles which are at
the basis of Landau's Fermi liquid theory. Near a quantum critical
point, these quasiparticles can be destroyed and non-Fermi
liquid behavior ensues. YbRh$_2$Si$_2$ is a prototypical correlated
metal as it exhibits quasiparticles formation, formation of Kondo
lattice coherence and quasiparticle destruction at a field-induced
quantum critical point. Here we show how, upon lowering the
temperature, the Kondo lattice coherence develops and finally
gives way to non-Fermi liquid electronic excitations. By measuring
the single-particle excitations through scanning tunneling
spectroscopy down to 0.3 K, we find the Kondo lattice peak
emerging below the Kondo temperature $T_{\rm K} \sim$ 25~K, yet
this peak displays a non-trivial temperature dependence with a
strong increase around 3.3 K. At the lowest temperature
and as a function of an external magnetic field, the width of
this peak is minimized in the quantum critical regime. Our
results provide a striking demonstration of the non-Fermi 
liquid electronic excitations in quantum critical metals, 
thereby elucidating the strange-metal phenomena that have been
ubiquitously observed in strongly correlated electron materials.}

Heavy fermion materials, {\it i.e.} intermetallics that contain
rare earths (REs) like Ce, Sm and Yb or actinides like U and Np,
are model systems to study strong electronic correlations
\cite{gre91,wir16}. The RE-derived localized $4f$ states can
give rise to local magnetic moments which typically order
(often antiferromagnetically) at sufficiently low temperature
as a  result of the inter-site Ruderman-Kittel-Kasuya-Yosida
interaction. In addition, the on-site Kondo effect causes a
hybridization between the 4$f$ and the conduction electrons,
which eventually screens the local moments by developing Kondo
spin-singlet many-body states. Hence, these two interactions
directly compete with each other and lead to different
(long-range magnetically ordered {\it vs.\ }paramagnetic
Fermi-liquid) ground states \cite{don77}. A zero-temperature
transition between the two states can be controlled through
doping, pressure or magnetic field $H$. A quantum critical point
(QCP) and concomitantly non-Fermi liquid properties ensue if
the phase transition is continuous at zero temperature
\cite{and87,col05,hvl07}.

Heavy fermion metals have been established as a canonical
setting for quantum criticality \cite{wir16}. How the Kondo
lattice coherence develops upon lowering the temperature,
{\it i.e.} the hierarchy of energy scales, is, however, still
a matter of debate. Intuitively, the coherence temperature
$T_{\rm coh}$ is set by the single-ion Kondo temperature
$T_{\rm K}$ of the lowest-lying crystal-field level \cite{ern11}
and can be further reduced by disorder \cite{pik12}, while
within another model $T_{\rm coh}$ can exceed $T_{\rm K}$
considerably \cite{yan08,yan16}. The latter might be related
to the influence of crystalline electric field effects
\cite{wir16,jang17}. Considerable experimental efforts have
recently been devoted to the study of the quantum critical
regime at relatively low temperatures. A key observation is
that quantum criticality induces a large entropy, suggesting
that it is linked with the Kondo effect. This raises an
important question as to how the onset of Kondo lattice
coherence at elevated temperatures connects with the
emergence of quantum criticality at low temperatures.

\begin{figure}[t]
\centering \includegraphics[width=8.6cm,clip=true]{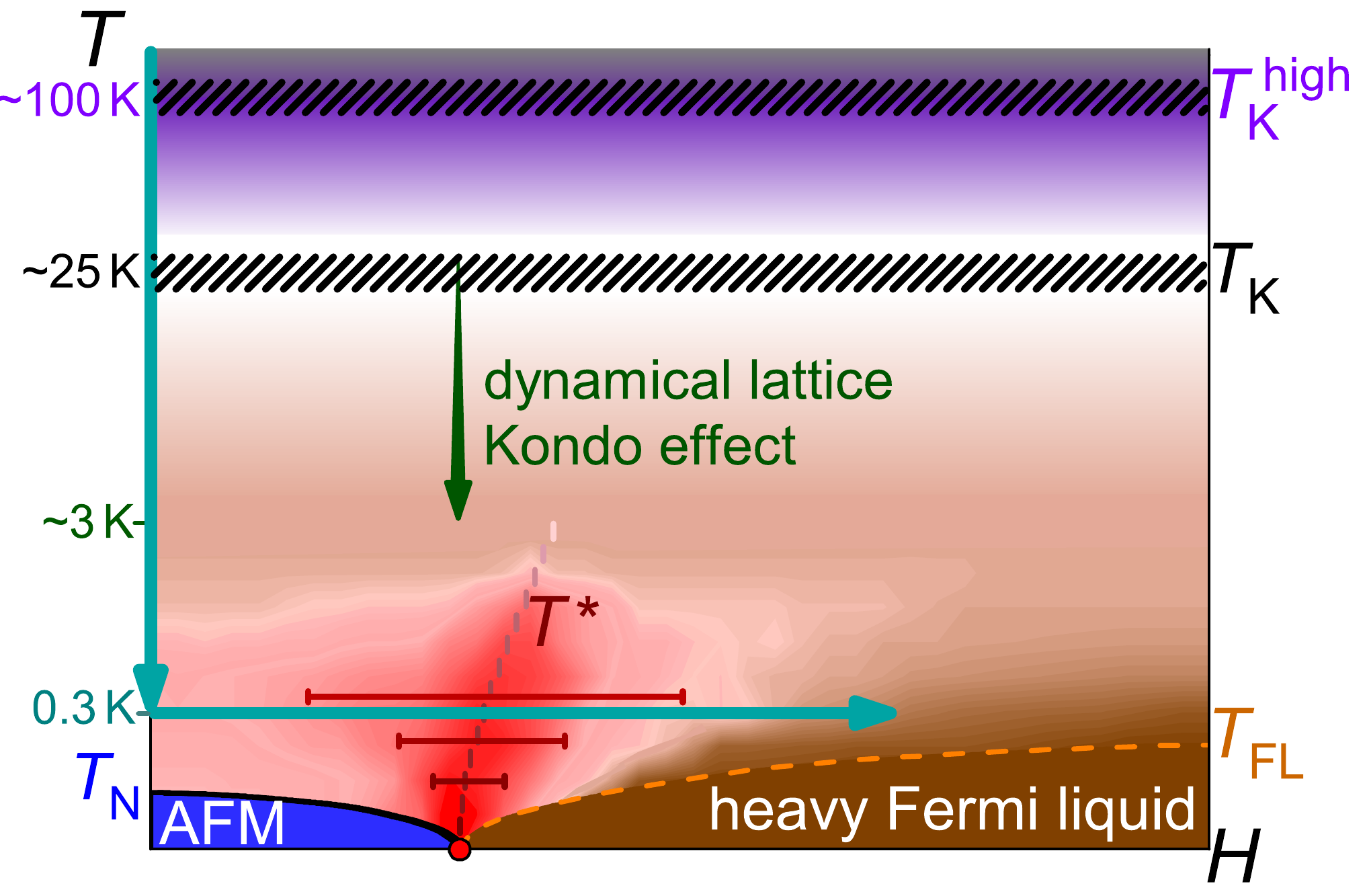}
\caption{\textbf{Phase diagram of YbRh$_2$Si$_2$.}
Schematic temperature--magnetic field phase diagram
according to the Kondo breakdown scenario.
The single-ion Kondo temperatures $T_{\rm K}^{\rm high}$ and
$T_{\rm K}$ involve all (purple shading) and the lowest-lying
(white) crystal electric field levels, respectively. The
lattice Kondo effect starts to develop around $T_{\rm K}$. The
Kondo-exchange interaction between the two types of spins,
respectively belonging to the local moments or the conduction
electrons, gives rise to dynamical correlations in the spin-singlet
channel. This dynamical lattice Kondo effect grows as temperature
is decreased. At large magnetic fields, lowering the temperature
eventually turns the dynamical lattice Kondo effect into a static
one; this occurs at temperatures below $T_{\rm FL}$, where the
system is in a heavy Fermi liquid state. For small magnetic fields
the lattice Kondo effect stays dynamical down to zero temperature.
Here, an antiferromagnetic (AFM) order develops below the N\'{e}el
temperature $T_{\rm N}$. $T^*(H)$ separates these two types of
behavior as the system evolves towards the respective ground
state. This scale, anchored by the QCP (red dot), marks the
finite-temperature signature of the Mott-type phase transition
at $T = 0$, with the thermal excitations of such a critical state
describing the quantum critical regime (red shading). The Fermi
surface jumps across such a QCP \cite{pas04,geg07,sven10}. The
light-green arrows indicate the parameters used in STS
measurements.} \label{phase} \end{figure}

The prototypical heavy fermion metal YbRh$_2$Si$_2$ shows an
antiferromagnetic ground state with a very low N\'eel
temperature, $T_{\rm N} = 70$ mK, and a QCP upon applying a
relatively small field $\mu_0 H_{\rm  N}= 0.66$ T parallel to the
tetragonal $c$-axis. Non-Fermi liquid behavior has been observed
in the quantum critical regime, extending up to temperatures of
about 0.5~K \cite{pas15}, depending on the physical quantity
that is measured as well as the degree of disorder \cite{jer03},
see $T$-$H$ phase diagram in Fig.\ \ref{phase}. Isothermal
magnetotransport and thermodynamic measurements at low
temperatures have provided evidence for the existence of an
additional low energy scale $T^*(H)$, which has been interpreted
as the finite-temperature manifestation of the critical destruction
of the lattice Kondo effect \cite{si14} and the concomitant
zero-temperature jump of the Fermi surface from large to small
across the QCP \cite{pas04,geg07,sven10}. Measurements of the
Gr\"{u}neisen and magnetic Gr\"{u}neisen ratio strongly support
this finding \cite{tok09,geg10}. Yet we note that alternative
scenarios have also been proposed \cite{abr12,woe15,miy14}.

Scanning tunneling spectroscopy (STS) measures {\it locally} the
density of states (DOS) \cite{fis07} through single-particle
excitations \cite{sch10,ern11,ayn12}. Spectra obtained at
temperatures $T \ge$ 4.6 K and $H =$ 0 revealed the successive
depopulation of the excited crystalline electric field (CEF)
states as the temperature is lowered, with essentially only
the lowest crystal-field Kramers doublet occupied at lowest
temperatures \cite{ern11}. The coupling between the localized 4$f$
electrons of this Kramers doublet and the conduction electrons
gives rise to periodic Kondo-singlet correlations which start to
develop below $T_{\rm coh}$. This coherence temperature is linked
to the effective single-ion Kondo temperature $T_K \approx 25$ K
extracted from bulk measurement \cite{koe08}. While these properties
conform to the traditional understanding of the high-temperature
behavior of the Kondo lattice \cite{geo,cos02}, the questions
remain open on how the Kondo coherence evolves further upon
lowering the temperature \cite{mo12,kum15,pas15} and in applied
field (light-green arrows in Fig.\ \ref{phase}) and,
importantly, how it connects with quantum criticality.

\subsection*{\sf\bfseries Results}
\begin{figure}[t]
\centering \includegraphics[width=8.6cm,clip=true]{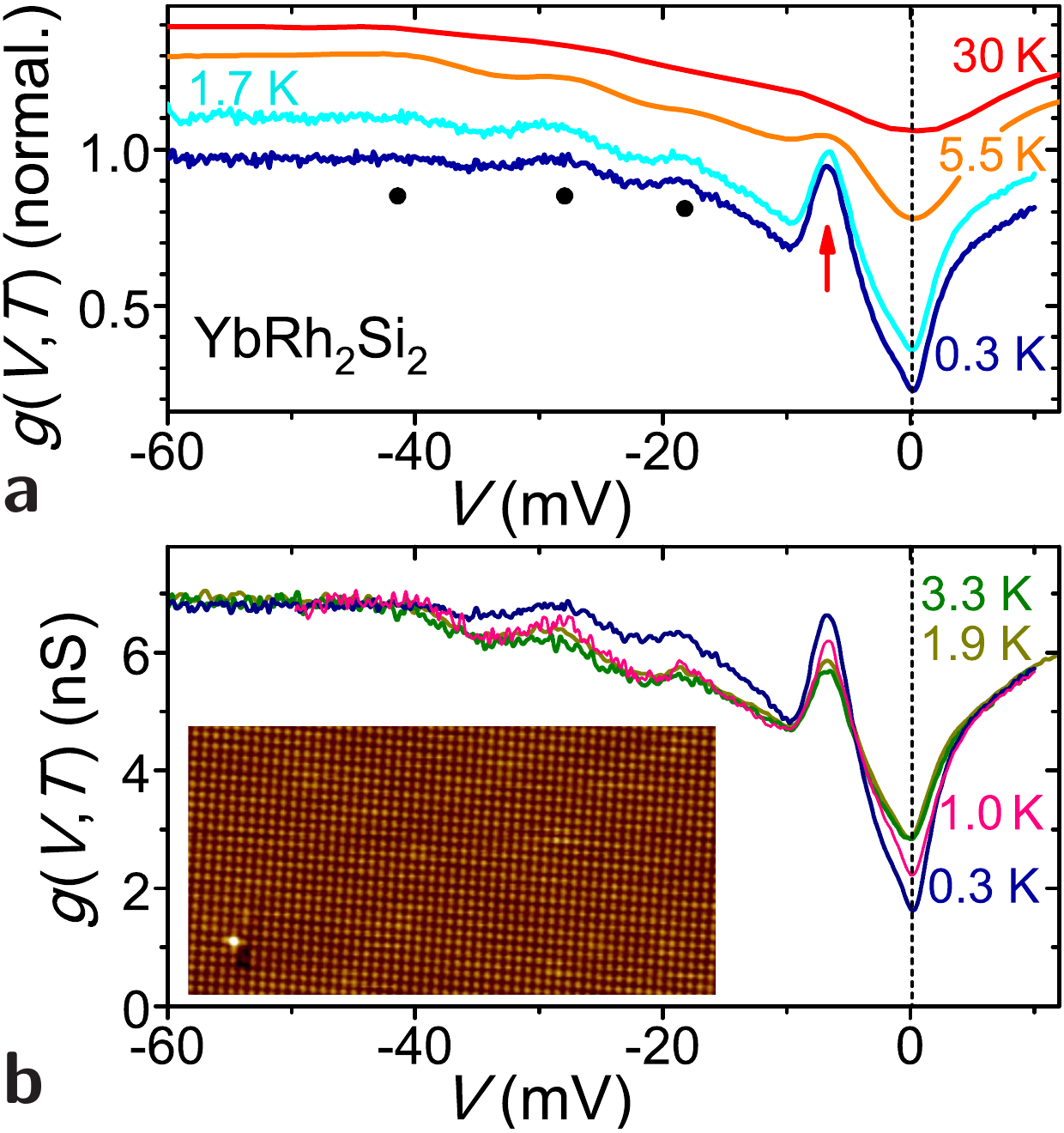}
\caption{\textbf{Tunneling spectroscopy on YbRh$_2$Si$_2$.}
{\bf a}, Tunneling conductance $g(V,T)$ normalized at $V = -80$
mV and obtained on different samples. Curves at 1.7 K, 5.5 K and
30 K are offset for clarity. The $-6$meV-peak evolves strongly at
low $T$ (red arrow). Black dots mark features resulting from
CEF splitting of the Yb $4f$ multiplet.
{\bf b}, $g(V,T)$-curves at selected low $T$ obtained on the
Si-terminated surface shown in the inset. Inset: Topography
indicating excellent surface quality ($20 \!\times\! 10$ nm$^2$,
$V \! =\!$ 100 mV, $I \! =\!$ 0.6 nA).}
\label{didv} \end{figure}

\noindent \textbf{Temperature evolution of tunneling spectra
down to 0.3 K}
Tunneling conductance curves d$I/$d$V = g(V,T)$ obtained over a
wide range of temperatures are presented in Fig.\ \ref{didv}a.
Both, the peaks due to CEF splitting of the Yb$^{3+}$
multiplet (marked by black dots in Fig.\ \ref{didv}a) and the
conductance dip at zero bias ($V=0$), result from single-ion
Kondo physics \cite{ern11}. Specifically the latter signifies
the hybridization between $4f$ and conduction electrons. The most
striking feature, however, is the evolution of the peak at about
$-6$ mV (red arrow in Fig.\ \ref{didv}a). Although this peak
appears to initially develop below 30 K, it clearly dominates
the spectra only for $T \lesssim 3.3$ K.

We now focus on this low temperature regime $T \leq $ 3.3~K,
Fig.\ \ref{didv}b. These data were obtained on the surface
shown in the inset where topography over an area of
$20 \times 10$ nm$^2$ is presented. Such a topography not only
attests the excellent sample quality but is also indicative of
Si termination (see Supplementary Note 1 and Supplementary
Figs.\ 1, 2). This termination is pivotal to our discussion as
it implies predominant tunneling into the conduction electron
states. A hint toward the origin of the $-6$~mV-peak comes from
renormalized band structure calculations \cite{zwi11}: a
partially developed hybridization gap is seen in the
quasiparticle DOS at slightly smaller energy. On the other
hand, a multi-level, finite-U non-crossing approximation (NCA)
without intersite correlations described our temperature dependent
\begin{figure}[t]
\centering \includegraphics[width=8.0cm,clip=true]{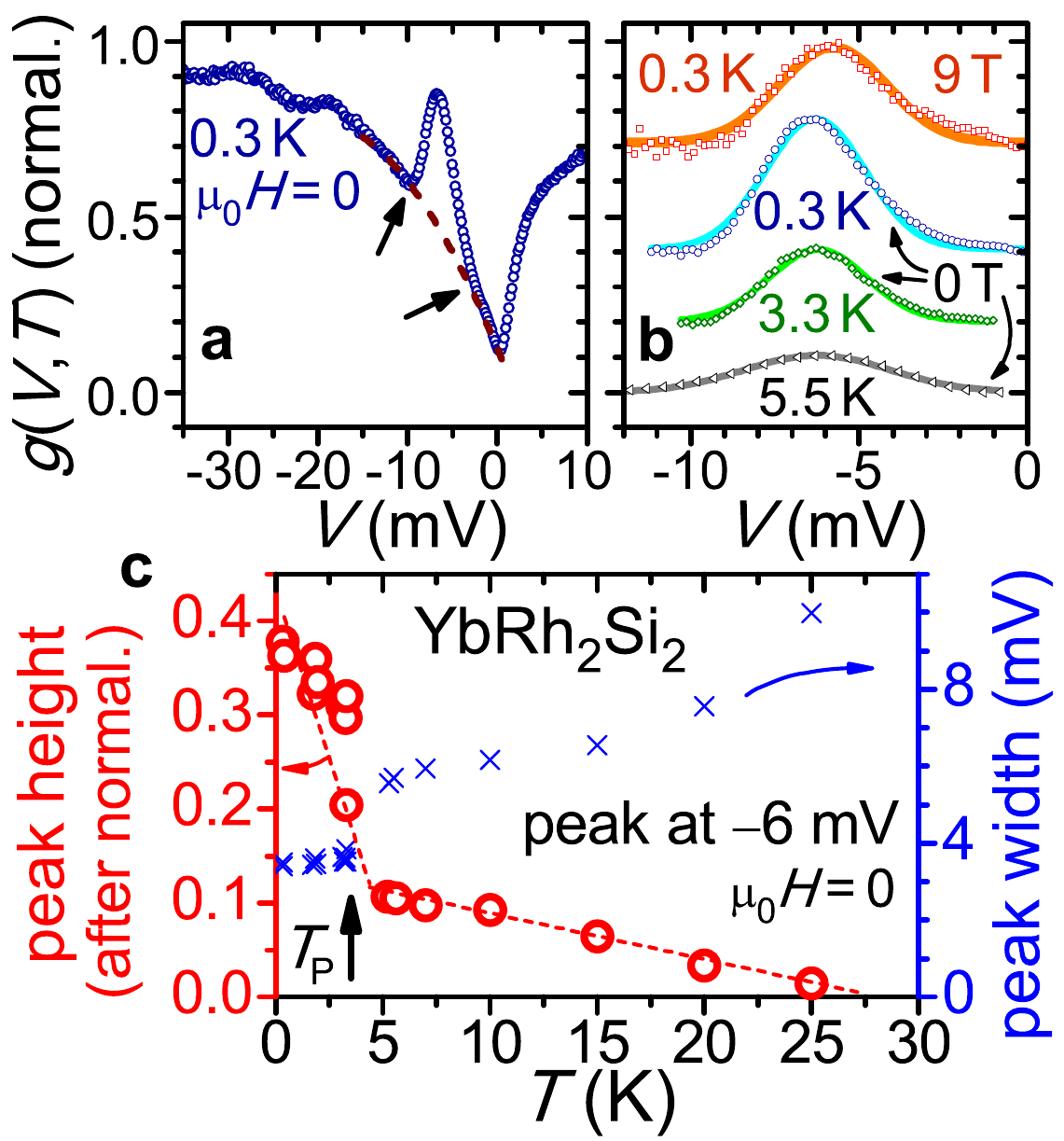}
\caption{\textbf{Analysis of the Kondo lattice peak.} {\bf a},
Tunneling conductance $g(V,T\!=\!0.3$ K) normalized to its value
at $V = -80$ mV, and parabola used for background subtraction
(dashed line). Arrows indicate onset of deviations between data
and parabola. {\bf b}, Examples of $g(V,T,H)$-data after
background subtraction (hollow markers, data sets at $T \leq
3.3$ K are offset). Data in zero as well as applied magnetic
field can be well described by Gaussians (lines). {\bf c},
Height and width (FWHM) of the peak at $-6$ mV after normalizing
all $g(V,T)$-curves at $-80$ mV. At $T_{\rm P}$, peak height
and width change significantly. Results from different samples
cause several markers to overlap. Dashed lines are guides to
the eye.} \label{peak}
\end{figure}
tunneling spectra away from the energy range of this peak
reasonably well \cite{ern11} but presented no indication for the
existence of a peak at $-$6 mV. Consequently, this peak must
result from the {\it lattice} Kondo effect, and will be
referred to as the Kondo lattice peak. The bulk nature of this
peak is supported by comparison to bulk transport measurements,
as discussed below.

An analysis of the Kondo lattice peak is impeded by the strongly
temperature-dependent zero-bias dip close by (see also
Supplementary Notes 2, 3 and Supplementary Figs. 3, 4). Data
$g(V,T \!\gtrsim \! 30\,$ K) for $-15$ mV $\leq V \leq -3$ mV can
be well approximated by a parabola and hence, we assume
a parabola to describe the background below the Kondo lattice peak
at low temperature, see the example of $T =$ 0.3 K in Fig.\
\ref{peak}a. There are finite energy ranges on both sides of the
peak feature allowing to fit a parabola, {\it cf.\ }arrows in Fig.\
\ref{peak}a. After background subtraction, each peak can be well
described by a Gaussian (lines in Fig.\ \ref{peak}b) from which its
height and width (full width at half maximum, FWHM) is extracted.
Note that the peak position in energy is independent of temperature.
Clearly both, the peak height and FWHM, exhibit a significant
change across \mbox{$T_{\rm P} \approx$ 3.3 K}, Fig.\ \ref{peak}c.

\begin{figure}[t]
\centering \includegraphics[width=8.2cm,clip=true]{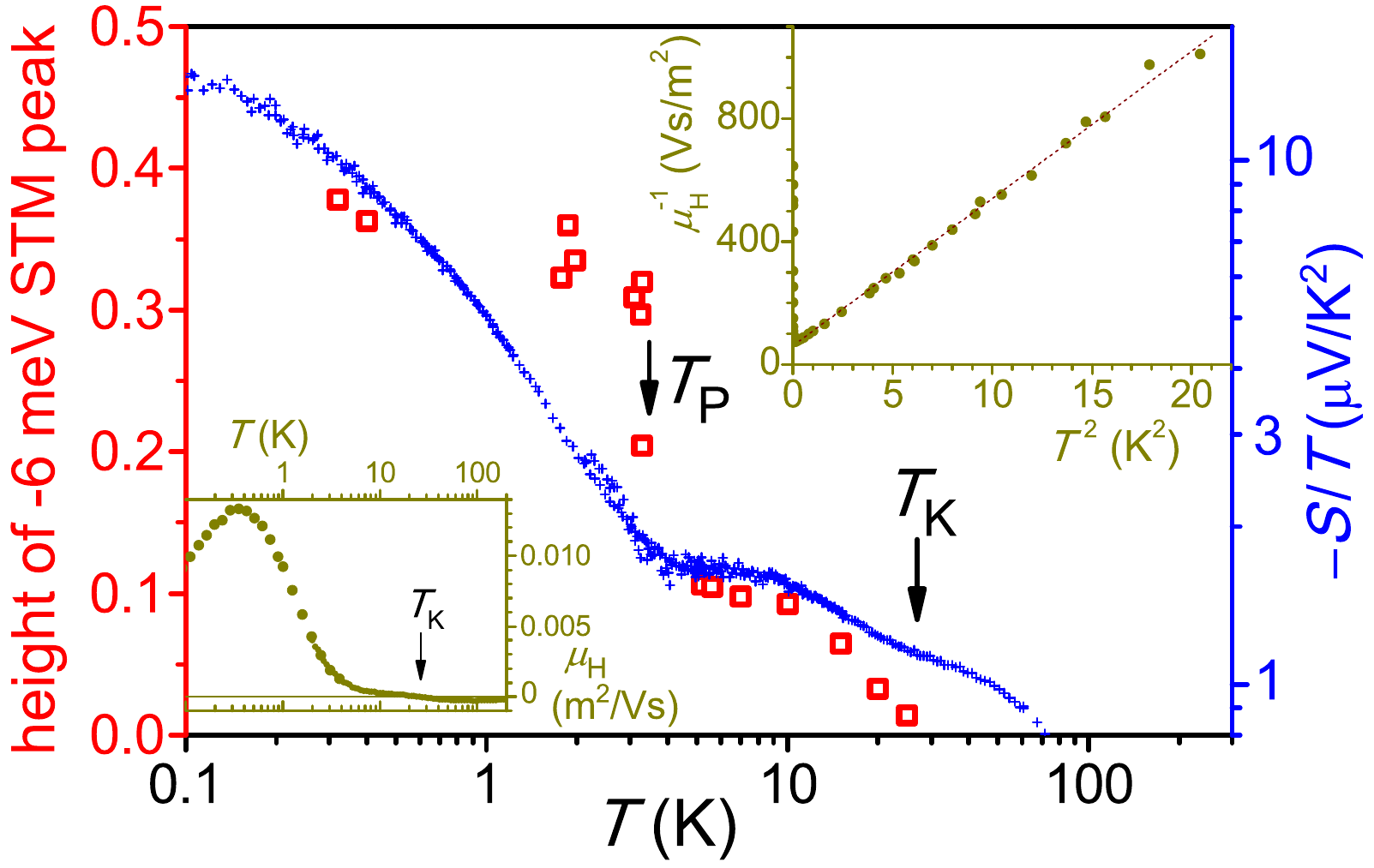}
\caption{\textbf{Development of dynamical lattice correlations}.
The height of the Kondo lattice peak is compared to thermopower $S$
divided by $T$, in dependence on $T$. Left inset: Hall mobility
$\mu_H$ vs. $T$. All three properties exhibit a strong upturn
below $T_{\rm P} \approx$ 3.3 K and saturation at lowest $T$.
Right inset: Power-law dependence of inverse Hall mobility
$1/\mu_H = \cot \theta_H$. The dashed straight line indicates
a $T^2$-dependence.} \label{compare} \end{figure}

While this temperature evolution of the single-particle
spectrum is surprising, it connects well with the features that
appear in bulk transport measurements
\cite{jer03,pas04,geg07,sven10,hart10,kim06}. Importantly, Fig.\
\ref{compare} shows that the thermopower divided by temperature,
$-S/T$, has a qualitatively similar temperature dependence as the
height of the STS Kondo lattice peak. Both display a plateau
around $3$ K, and a subsequent strong increase upon lowering the
temperature. For the thermopower, the plateau signifies the
formation of some kind of Fermi liquid, {\it i.e.} prevailing
lattice Kondo correlations, while the regime below about $3$ K
reflects non-Fermi liquid behavior associated with quantum
criticality \cite{hart10}. This suggests that, in the same
temperature regime, the growth of the STS Kondo lattice peak
height as well as the concomitant drop of peak width (Fig.\
\ref{peak}c) also capture the quantum critical behavior. In
this picture, quantum criticality arises when there is
sufficient buildup of dynamical lattice Kondo correlations
(see Supplementary Note 4), or conversion of the 4$f$ electron
spins into quasiparticle-like, but still incoherent excitations.

To illustrate this point further, the Hall mobility $\mu_H =
R_H / \rho_{xx}$ as a function of temperature is also plotted in
Fig.\ \ref{compare}, left inset. In the regime where the anomalous
Hall effect dominates, this quantity has been considered as
capturing the buildup of the on-site Kondo resonance \cite{col85}.
It is striking that the Hall mobility also shows a strong
increase upon lowering the temperature. Interestingly, the Hall
mobility does not show any plateau near 3 K, and neither does
the resistivity nor the Sommerfeld coefficient as a function of
temperature \cite{jer03}. This implies that $T_{\rm P} \approx$
3.3 K is not an ordinary Fermi liquid scale. Instead, an
additional broad peak occurs in the Hall coefficient at around
0.7 K (see Supplementary Note 5 and Supplementary Fig.\ 6). The
connection between the growth of the Hall mobility with quantum
criticality becomes evident when we analyze its inverse $1/\mu_H
=  \rho_{xx} /R_H$, which is equivalent to the cotangent of the
Hall angle, $\cot \theta_H$, as a function of temperature.
\begin{figure}[t]
\centering \includegraphics[width=7.6cm,clip=true]{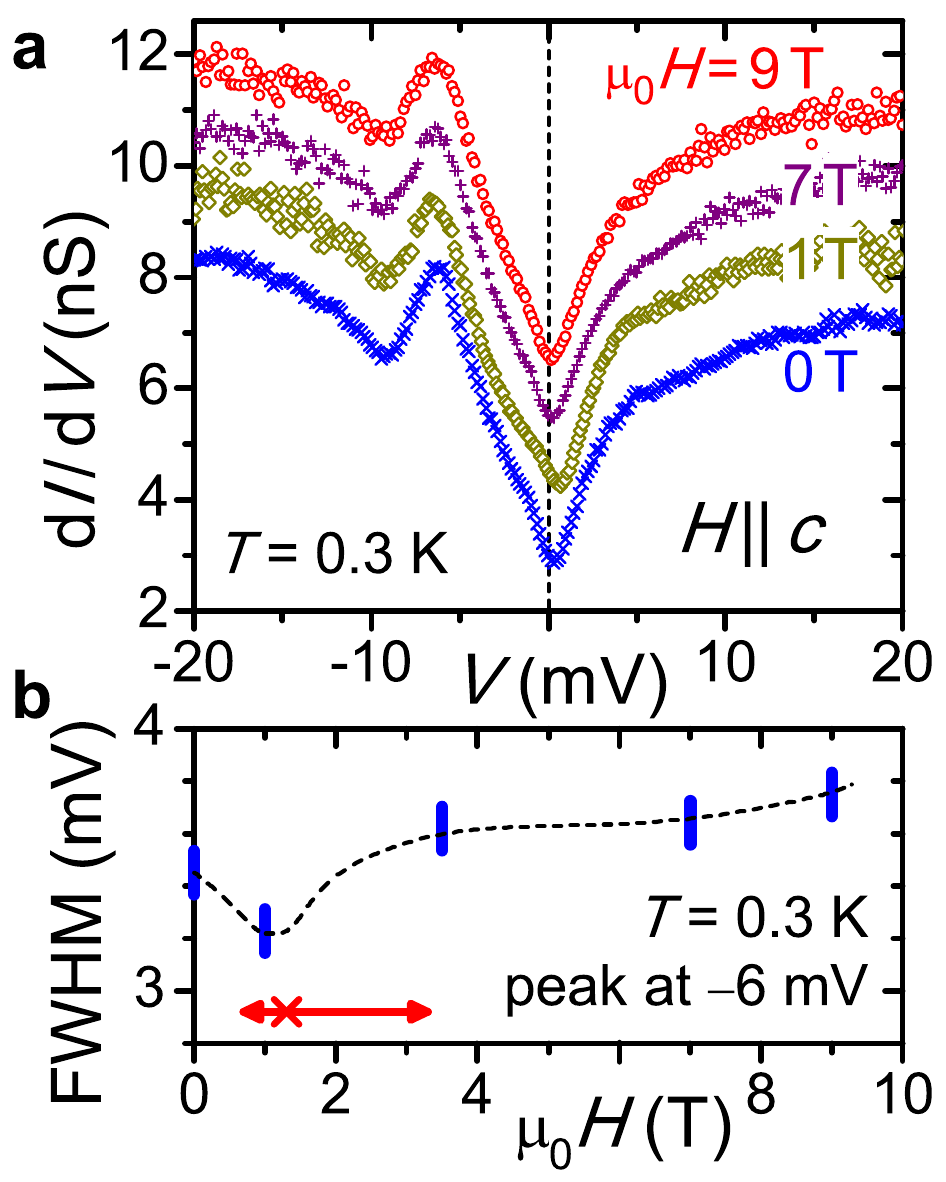}
\caption{\textbf{Spectroscopy in applied field.} {\bf a}, Tunneling
conductance $g(V,H,T\!=\!0.3$ K) measured at different magnetic
fields applied parallel to the magnetically hard $c$ axis.
Curves are offset for clarity. {\bf b}, FWHM of the Kondo lattice
peak for different magnetic fields at $T = 0.3$ K. At this
temperature and field orientation, the energy scale $T^*$ ({\it
cf.} Fig.\ \ref{phase}) is located at a field of about 1.3 T (red
cross, Refs. \cite{pas04,geg07}), approximately where a minimum is
observed in the peak width. The red arrow indicates the FWHM of
the Hall crossover at $T =$ 0.3 K, Ref.\ \cite{sven10}. Blue
marker heights correspond to the errors of the Gaussian fits
(Figs.\ \ref{peak}a,b) and differences between samples, the line
is a guide to the eye.} \label{field} \end{figure}
$1/\mu_H$ obeys a power-law behavior, $1/\mu_H \sim T^2$, for
0.5~K $\lesssim T \lesssim$ 5~K ({\it cf.\ }Fig.\ \ref{compare},
right inset); such a behavior, together with a $T$-linear
electrical resistivity $\rho(T)$, has also been observed,
{\it e.g.,} in the cuprate high-$T_{\rm c}$ superconductors
\cite{ong91}.\\

\noindent \textbf{Evolution of tunneling spectra in magnetic
fields}
To search for more direct STS evidence for quantum criticality
in the $H$--$T$ phase diagram of YbRh$_2$Si$_2$, the system
was tuned by a magnetic field at $T = 0.3$ K. At this $T$, the
large body of thermodynamic, magnetotransport and magnetic
measurements (see \cite{pas15} and references therein) did
{\it not} give any indication for any other feature except the
$T^*(B)$-line, {\it cf.} Fig.\ \ref{phase}, a fact that is
pivotal for the discussion below. Some $g(V,H,T\!=\!
0.3\,$K)-curves are presented in Fig.\ \ref{field}a. No major
change in the overall shape of the spectra with magnetic field
is observed. The Kondo lattice peak can again be described
by a Gaussian, see example of $\mu_0 H = 9$ T included in Fig.\
\ref{peak}b. The FWHM dependence on $H$ is shown in Fig.\
\ref{field}b. We note that the FWHM at low $T$ varies very
little between different spectra, and even different samples,
{\it i.e.} $< 4$\% (see also Fig.\ \ref{peak}c where several
data points of the FWHM fall on top of each other). This is
taken as the error of FWHM and determines the marker size in
Fig.\ \ref{field}b. Moreover, a comparison between the data and
the Gaussian fit in Fig.\ \ref{peak}b reveals an only slightly
enhanced noise of $g(V,H,T\!=\! 0.3\,$K) at $\mu_0 H = 9$ T
compared to zero field. Consequently, the trend displayed in
Fig.\ \ref{field}b appears genuine.

At a field of $\mu_0 H = 1$ T, the Kondo lattice peak FWHM
exhibits a minimum, with a reduction of about 15\% of its
high-field value. This field is approximately of the value
$\mu_0 H^* \approx 1.3$ T at which the Hall crossover
\cite{pas04,geg07,sven10} takes place at $T = 0.3$~K for
$H||c$ (red cross in Fig.\ \ref{field}b, see Supplementary Note
6). The range in magnetic field over which the Hall crossover
is observed \cite{sven10} and hence, changes in $g(V,H,T\!=\!
0.3\,$K) are to be expected, is indicated by a red arrow in
Fig.\ \ref{field}b. Clearly, it agrees well with the width of
the drop in peak width {\it vs.\ }$H$ we observe at $T = 0.3$
K. This drop is expected to further increase and sharpen upon
cooling, {\it cf.\ }Fig.\ \ref{phase}). This is consistent
with a critical slowing down concluded from isothermal
magnetotransport (Hall coefficient, $R_H$, and
magnetoresistance, $\rho_{xx}$) measurements
\cite{pas04,sven10}, which reveal thermally broadened jumps at
$H^*(T)$. They indicate the finite temperature remnant of a
reconstruction of the Fermi surface at the field-induced QCP,
{\it i.e.} at $T = 0$, where the quasiparticle weight vanishes
\cite{pfau12}, {\it cf.\ }Supplementary Note 7. Data
from specific-heat measurements on YbRh$_2$Si$_2$ in magnetic
field \cite{oes08} confirm this scenario ({\it cf.\
}Supplementary Fig.\ 7). They yield a relative change of the
Sommerfeld coefficient between critical ($H^*$) and elevated
fields of order 30\% at $T=0.3$ K, if scaled for the relevant
field orientation. We speculate the larger change in
Sommerfeld coefficient compared to the drop in FWHM of the STS
Kondo lattice peak (about 15\% in Fig.\ \ref{field}b) is related
to the fact that heat capacity integrates over the whole
Brillouin zone while STS is a more directional measurement.
For a surface along the $a$-$b$ plane (Fig.\ \ref{didv}),
tunneling along the $c$-direction is the most relevant, yet
hybridization of the Yb CEF ground state orbitals is anisotropic
\cite{zwi11}, mostly with the Rh $4d_{x^2 - y^2}$. Moreover, the
specific heat is determined by the inverse of the quasiparticle
weight while the width reflects a critical slowing down. In
other words, the specific heat captures the real part of the
self-energy, while the width reflects the imaginary part. It
should also be noted that even at a temperature as low as
$\sim$0.5 K, the Hall crossover reaches all the way to $B \! =
\! 0$ \cite{pas15}. Because of the dominating contribution of
the large Fermi surface to the quantum-critical fluctuations
\cite{sen08}, the peak width in STS at $B \! = \! 0$ should be
close to the one extrapolated from higher fields, where a large
Fermi surface constitutes the heavy Fermi liquid. Upon cooling
to below about 0.5~K, this contribution of the large Fermi
surface at $B\! = \! 0$ is expected to decrease \cite{pas15}.
Indeed, at $T\! =\! 0.3$ K the peak width at $B \! = \! 0$
appears to be slightly reduced compared to an extrapolation
from high fields, Fig.\ \ref{field}b. We also note that Lifshitz
transitions and Zeeman splitting can be ruled out as origins
for the drop of the peaks' FWHM (see Supplementary Note 5).

\subsection*{\sf\bfseries Discussion}
Our STS studies here have revealed two important new insights.
One is that the development of the dynamical lattice Kondo
correlations in a stoichiometric material such as YbRh$_2$Si$_2$,
while setting in at $T_{\rm coh} \approx T_{\rm K}$, extends to
considerably lower temperatures and dominate the material's
properties only at much lower temperatures (see Supplementary
Note 4). In the case of YbRh$_2$Si$_2$, the STS Kondo lattice
peak height and thermopower coefficient do not indicate dominant
lattice Kondo correlations before the temperature has reached
$T_{\rm P} \sim 0.1 \!\cdot\! T_{\rm coh}$. Moreover, the
conductance minimum at zero bias, which has been shown to capture
the onset of the on-site Kondo ({\it i.e.} hybridization) effect
at high temperatures \cite{ern11}, continues to deepen down to
the lowest measured temperature as is seen in Figs.\
\ref{didv}a, b. Conversely, the strengthening of the lattice
Kondo coherence only at much below $T_{\rm K}$ implies
that the on-site Kondo effect dominates many thermodynamic and
transport properties at around $T_{\rm coh}$ in YbRh$_2$Si$_2$,
and gives way to the lattice Kondo correlations only slowly
upon reducing the temperature. Such a persistence of this
distinct signature of the single-ion Kondo effect down to
temperatures substantially below $T_{\rm coh}$ is consistent
with observations based on different transport \cite{col85,sun13}
and thermodynamic \cite{pie05,pik12} properties of several other
heavy-fermion metals. This also provides a natural explanation
to the applicability of single-ion-based descriptions to
temperatures well below $T_{\rm K}$ even though they neglect
lattice Kondo coherence effects \cite{col85,ern11}.

The second lesson concerns the link between the development of
the dynamical lattice Kondo correlations and quantum
criticality. As a function of temperature, our measurements of
the height and width of the Kondo lattice peak suggest that,
in order for the quantum criticality to set in, the lattice
Kondo correlations first have to develop sufficiently upon
lowering the temperature through, and well below, $T_{coh}
\cong 30$~K. More specifically, as the temperature is lowered
through $T_{\rm coh}$, both the Kondo lattice peak height and
the thermopower coefficient first reach a plateau at around 3
K signifying well developed lattice Kondo correlations. It is
against this backdrop that the Kondo lattice peak height markedly
increases below $T_{\rm P}$. This manifests quantum criticality
at the level of the single-particle spectrum, which goes
considerably beyond the quantum critical behavior seen in the
divergent Sommerfeld coefficient of the electronic specific heat
and the linear-in-$T$ electrical resistivity\cite{jer03}. This
signature of the quantum criticality at the single-particle level
is complemented by the isothermal behavior of the Kondo lattice
peak with respect to the control parameter, the magnetic field,
at the lowest measured temperature, $T \approx 0.3$ K. The
width displays a minimum, ascribed to a critical slowing
down associated with the Kondo destruction, at the field
$H^*(T\! = \! 0.3$ K) of similar value as determined by
isothermal measurements of transport and thermodynamic
quantities \cite{pas04,geg07,sven10}.

To put these findings into perspective, our comparison
between STS and magnetotransport measurements show that the
development of the lattice Kondo correlations is the prerequisite
for the localization-delocalization transition associated with
the unconventional quantum critical point. This reveals the
Mott physics in quantum critical heavy fermion metals at the
single-particle level. As such, the insights gained in our study
will likely be relevant to the non-Fermi liquid phenomena in a
broad range of other strongly correlated metals such as the
high-$T_c$ cuprates and the organic charge-transfer salts, which
are typically in proximity to Mott insulating states and in
which quantum criticality is often observed
\cite{lee06,ram15,oike15}.\\

\section*{Methods}
High-quality single crystals of YbRh$_2$Si$_2$ were grown by an
indium-flux method; they grow as thin platelets with a height of
0.2--0.4 mm along the crystallographic $c$-direction (see also
Supplementary Note 6). Crystalline quality and orientation of the
single crystals were confirmed by x-ray and Laue investigations,
respectively. The residual resistivity $\rho_0$ of the 6 samples
investigated here ranged between 0.5 $\mu\Omega \,$cm and $0.9
\;\mu\Omega \,$cm with no apparent differences in their
spectroscopic results. The samples were cleaved {\it in situ}
perpendicular to the crystallographic $c$ direction at
temperatures $\lesssim 20$ K. Subsequent to cleaving, the samples
were constantly kept under ultra-high vacuum (UHV) conditions
and did not exhibit any sign of surface degradation for at least
several months, as indicated by STM re-investigation.

STM and STS was conducted (using a cryogenic STM made by Omicron
Nanotechnology) at temperatures between 0.3 -- 6 K, in magnetic
fields $\mu_0 H \leq 12$ T (applied parallel to the
crystallographic $c$ direction) and under UHV conditions ($p < 2
\cdot 10^{-9}$ Pa). For the temperature range 4.6 K $\leq T \leq$
120 K a second UHV STM (LT-STM) was utilized ($p \leq 3 \cdot
10^{-9}$ Pa).

\section*{Acknowledgements}
We sincerely thank U.\ Stockert for help with the experiments
and discussions, as well as J.C.\ S\'{e}amus Davis for
discussions. Work was partly supported by the German
Research Foundation through DFG Research Unit 960 and through
DFG grant KR3831/4-1. Work at Rice University has been supported
by the NSF Grant DMR-1611392 and the Robert A.\ Welch Foundation
Grant No.\ C-1411. Q.S.\ graciously acknowledges the support of
the Alexander von Humboldt Foundation and the hospitality of the
Karlsruhe Institute of Technology. S.~Kirchner acknowledges
support by the National Key R\&D Program of the MOST of China
(No.\ 2016YFA0300202 and No.\ 2017YFA0303100) and the National
Science Foundation of China (No.\ 11474250 and No.\ 11774307).


\begin{thebibliography}{10}
\expandafter\ifx\csname url\endcsname\relax
  \def\url#1{\texttt{#1}}\fi
\expandafter\ifx\csname urlprefix\endcsname\relax\def\urlprefix{URL }\fi
\providecommand{\bibinfo}[2]{#2}
\providecommand{\eprint}[2][]{\url{#2}}

\bibitem{gre91}
\bibinfo{author}{Grewe, N.} \& \bibinfo{author}{Steglich, F.}
\newblock \bibinfo{title}{Heavy fermion metals}.
\newblock In \bibinfo{editor}{{Gschneidner, Jr.}, K.~A.} \&
  \bibinfo{editor}{Eyring, L.} (eds.) \emph{\bibinfo{booktitle}{Handbook on the
  Physics and Chemistry of Rare Earths}}, vol.~\bibinfo{volume}{14},
  \bibinfo{pages}{343--474} (\bibinfo{publisher}{Elsevier Amsterdam},
  \bibinfo{year}{1991}).

\bibitem{wir16}
\bibinfo{author}{Wirth, S.} \& \bibinfo{author}{Steglich, F.}
\newblock \bibinfo{title}{Exploring heavy fermions from macroscopic to
  microscopic length scales}.
\newblock \emph{\bibinfo{journal}{Nature Rev.\ Mater.}}
  \textbf{\bibinfo{volume}{1}}, \bibinfo{pages}{16051} (\bibinfo{year}{2016}).

\bibitem{don77}
\bibinfo{author}{Doniach, S.}
\newblock \bibinfo{title}{Kondo lattice and weak antiferromagnetism}.
\newblock \emph{\bibinfo{journal}{Physica B}} \textbf{\bibinfo{volume}{91}},
  \bibinfo{pages}{231--234} (\bibinfo{year}{1977}).

\bibitem{and87}
\bibinfo{author}{Anderson, P.}
\newblock \bibinfo{title}{The resonating valence bond state in la$_2$cuo$_4$
  and superconductivity?}
\newblock \emph{\bibinfo{journal}{Science}} \textbf{\bibinfo{volume}{235}},
  \bibinfo{pages}{1196--1198} (\bibinfo{year}{1987}).

\bibitem{col05}
\bibinfo{author}{Coleman, P.} \& \bibinfo{author}{Schofield, A.~J.}
\newblock \bibinfo{title}{Quantum criticality}.
\newblock \emph{\bibinfo{journal}{Nature}} \textbf{\bibinfo{volume}{433}},
  \bibinfo{pages}{226--229} (\bibinfo{year}{2005}).

\bibitem{hvl07}
\bibinfo{author}{{von L{\"o}hneysen}, H.}, \bibinfo{author}{Rosch, A.},
  \bibinfo{author}{Vojta, M.} \& \bibinfo{author}{W{\"o}lfle, P.}
\newblock \bibinfo{title}{Fermi-liquid instabilities at magnetic quantum phase
  transitions}.
\newblock \emph{\bibinfo{journal}{Rev.\ Mod.\ Phys.}}
  \textbf{\bibinfo{volume}{79}}, \bibinfo{pages}{1015--1075}
  (\bibinfo{year}{2007}).

\bibitem{pik12}
\bibinfo{author}{Pikul, A.} \emph{et~al.}
\newblock \bibinfo{title}{Single-ion {K}ondo scaling of the coherent {F}ermi
  liquid regime in {C}e$_{1-x}${L}a$_x${N}i$_2${G}e$_2$}.
\newblock \emph{\bibinfo{journal}{Phys.\ Rev.\ Lett.}}
  \textbf{\bibinfo{volume}{108}}, \bibinfo{pages}{066405}
  (\bibinfo{year}{2012}).

\bibitem{yan08}
\bibinfo{author}{Yang, Y.}, \bibinfo{author}{Fisk, Z.}, \bibinfo{author}{Lee,
  H.-O.}, \bibinfo{author}{Thompson, J.~D.} \& \bibinfo{author}{Pines, D.}
\newblock \bibinfo{title}{Scaling the {K}ondo lattice}.
\newblock \emph{\bibinfo{journal}{Nature}} \textbf{\bibinfo{volume}{454}},
  \bibinfo{pages}{611--613} (\bibinfo{year}{2008}).

\bibitem{yan16}
\bibinfo{author}{Yang, Y.}
\newblock \bibinfo{title}{Two-fluid model for heavy electron physics}.
\newblock \emph{\bibinfo{journal}{Rep.\ Prog.\ Phys.}}
  \textbf{\bibinfo{volume}{79}}, \bibinfo{pages}{074501}
  (\bibinfo{year}{2016}).

\bibitem{jang17}
\bibinfo{author}{Jang, S.} \emph{et~al.}
\newblock \bibinfo{title}{Evolution of the {K}ondo lattice electronic structure
  above the transport coherence temperature} (\bibinfo{year}{2017}).
\newblock \bibinfo{note}{{a}rXiv:1704.08247}.

\bibitem{pas15}
\bibinfo{author}{Paschen, S.} \emph{et~al.}
\newblock \bibinfo{title}{Kondo destruction in heavy fermion quantum
  criticality and the photoemission spectrum of {Y}b{R}h$_2${S}i$_2$}.
\newblock \emph{\bibinfo{journal}{J.\ Magn.\ Magn.\ Mater.}}
  \textbf{\bibinfo{volume}{400}}, \bibinfo{pages}{17--22}
  (\bibinfo{year}{2016}).

\bibitem{jer03}
\bibinfo{author}{Custers, J.} \emph{et~al.}
\newblock \bibinfo{title}{The break-up of heavy electrons at a quantum critical
  point}.
\newblock \emph{\bibinfo{journal}{Nature}} \textbf{\bibinfo{volume}{424}},
  \bibinfo{pages}{524--527} (\bibinfo{year}{2003}).

\bibitem{si14}
\bibinfo{author}{Si, Q.} \emph{et~al.}
\newblock \bibinfo{title}{Kondo destruction and quantum criticality in {K}ondo
  lattice systems}.
\newblock \emph{\bibinfo{journal}{J.\ Phys.\ Soc.\ Jpn.}}
  \textbf{\bibinfo{volume}{83}}, \bibinfo{pages}{061005}
  (\bibinfo{year}{2014}).

\bibitem{pas04}
\bibinfo{author}{Paschen, S.} \emph{et~al.}
\newblock \bibinfo{title}{Hall-effect evolution across a heavy-fermion quantum
  critical point}.
\newblock \emph{\bibinfo{journal}{Nature}} \textbf{\bibinfo{volume}{432}},
  \bibinfo{pages}{881--885} (\bibinfo{year}{2004}).

\bibitem{geg07}
\bibinfo{author}{Gegenwart, P.} \emph{et~al.}
\newblock \bibinfo{title}{Multiple energy scales at a quantum critical point}.
\newblock \emph{\bibinfo{journal}{Science}} \textbf{\bibinfo{volume}{315}},
  \bibinfo{pages}{969--971} (\bibinfo{year}{2007}).

\bibitem{sven10}
\bibinfo{author}{Friedemann, S.} \emph{et~al.}
\newblock \bibinfo{title}{Fermi-surface collapse and dynamical scaling near a
  quantum-critical point}.
\newblock \emph{\bibinfo{journal}{Proc.\ Natl.\ Acad.\ Sci.\ USA}}
  \textbf{\bibinfo{volume}{107}}, \bibinfo{pages}{14547--14551}
  (\bibinfo{year}{2010}).

\bibitem{tok09}
\bibinfo{author}{Tokiwa, Y.}, \bibinfo{author}{Radu, T.},
  \bibinfo{author}{Geibel, C.}, \bibinfo{author}{Steglich, F.} \&
  \bibinfo{author}{Gegenwart, P.}
\newblock \bibinfo{title}{Divergence of the magnetic {G}r{\"u}neisen ratio at
  the field-induced quantum critical point in {Y}b{R}h$_2${S}i$_2$}.
\newblock \emph{\bibinfo{journal}{Phys.\ Rev.\ Lett.}}
  \textbf{\bibinfo{volume}{102}}, \bibinfo{pages}{066401}
  (\bibinfo{year}{2009}).

\bibitem{geg10}
\bibinfo{author}{Gegenwart, P.} \emph{et~al.}
\newblock \bibinfo{title}{Divergence of the {G}r{\"u}neisen parameter and
  magnetocaloric effect at heavy fermion quantum critical points}.
\newblock \emph{\bibinfo{journal}{J.\ Low Temp.\ Phys.}}
  \textbf{\bibinfo{volume}{161}}, \bibinfo{pages}{117--133}
  (\bibinfo{year}{2010}).

\bibitem{abr12}
\bibinfo{author}{Abrahams, E.} \& \bibinfo{author}{W{\"o}lfle, P.}
\newblock \bibinfo{title}{Critical quasiparticle theory applied to heavy
  fermion metals near an antiferromagnetic quantum phase transition}.
\newblock \emph{\bibinfo{journal}{Proc.\ Natl.\ Acad.\ Sci.\ USA}}
  \textbf{\bibinfo{volume}{109}}, \bibinfo{pages}{3238--3242}
  (\bibinfo{year}{2012}).

\bibitem{woe15}
\bibinfo{author}{W{\"o}lfle, P.} \& \bibinfo{author}{Abrahams, E.}
\newblock \bibinfo{title}{Spin-flip scattering of critical quasiparticles and
  the phase diagram of {Y}b{R}h$_2${S}i$_2$}.
\newblock \emph{\bibinfo{journal}{Phys.\ Rev.\ B}}
  \textbf{\bibinfo{volume}{92}}, \bibinfo{pages}{155111}
  (\bibinfo{year}{2015}).

\bibitem{miy14}
\bibinfo{author}{Miyake, K.} \& \bibinfo{author}{Watanabe, S.}
\newblock \bibinfo{title}{Unconventional quantum criticality due to critical
  valence transition}.
\newblock \emph{\bibinfo{journal}{J.\ Phys.\ Soc.\ Jpn.}}
  \textbf{\bibinfo{volume}{83}}, \bibinfo{pages}{061006}
  (\bibinfo{year}{2014}).

\bibitem{fis07}
\bibinfo{author}{Fischer, {\O}.}, \bibinfo{author}{Kugler, M.},
  \bibinfo{author}{Maggio-{A}prile, I.}, \bibinfo{author}{Berthod, C.} \&
  \bibinfo{author}{Renner, C.}
\newblock \bibinfo{title}{Scanning tunneling spectroscopy of high-temperature
  superconductors}.
\newblock \emph{\bibinfo{journal}{Rev.\ Mod.\ Phys.}}
  \textbf{\bibinfo{volume}{79}}, \bibinfo{pages}{353--419}
  (\bibinfo{year}{2007}).

\bibitem{sch10}
\bibinfo{author}{Schmidt, A.~R.} \emph{et~al.}
\newblock \bibinfo{title}{Imaging the {F}ano lattice to `hidden order'
  transition in {U}{R}u$_2${S}i$_2$}.
\newblock \emph{\bibinfo{journal}{Nature}} \textbf{\bibinfo{volume}{465}},
  \bibinfo{pages}{570--576} (\bibinfo{year}{2010}).

\bibitem{ern11}
\bibinfo{author}{Ernst, S.} \emph{et~al.}
\newblock \bibinfo{title}{Emerging local {K}ondo screening and spatial
  coherence in the heavy-fermion metal {Y}b{R}h$_2${S}i$_2$}.
\newblock \emph{\bibinfo{journal}{Nature}} \textbf{\bibinfo{volume}{474}},
  \bibinfo{pages}{362--366} (\bibinfo{year}{2011}).

\bibitem{ayn12}
\bibinfo{author}{Aynajian, P.} \emph{et~al.}
\newblock \bibinfo{title}{Visualizing heavy fermions emerging in a quantum
  critical {K}ondo lattice}.
\newblock \emph{\bibinfo{journal}{Nature}} \textbf{\bibinfo{volume}{486}},
  \bibinfo{pages}{201--206} (\bibinfo{year}{2012}).

\bibitem{koe08}
\bibinfo{author}{K{\"o}hler, U.}, \bibinfo{author}{Oeschler, N.},
  \bibinfo{author}{Steglich, F.}, \bibinfo{author}{Maquilon, S.} \&
  \bibinfo{author}{Fisk, Z.}
\newblock \bibinfo{title}{Energy scales of {L}u$_{1-x}${Y}b$_x${R}h$_2${S}i$_2$
  by means of thermopower investigations}.
\newblock \emph{\bibinfo{journal}{Phys.\ Rev.\ B}}
  \textbf{\bibinfo{volume}{77}}, \bibinfo{pages}{104412}
  (\bibinfo{year}{2008}).

\bibitem{geo}
\bibinfo{author}{Burdin, S.}, \bibinfo{author}{Georges, A.} \&
  \bibinfo{author}{Grempel, D.~R.}
\newblock \bibinfo{title}{Coherence scale of the {K}ondo lattice}.
\newblock \emph{\bibinfo{journal}{Phys.\ Rev.\ Lett.}}
  \textbf{\bibinfo{volume}{85}}, \bibinfo{pages}{1048--1051}
  (\bibinfo{year}{2000}).

\bibitem{cos02}
\bibinfo{author}{Costi, T.~A.} \& \bibinfo{author}{Manini, N.}
\newblock \bibinfo{title}{Low-energy scales and temperature-dependent
  photoemission of heavy fermions}.
\newblock \emph{\bibinfo{journal}{J.\ Low Temp.\ Phys.}}
  \textbf{\bibinfo{volume}{126}}, \bibinfo{pages}{835--866}
  (\bibinfo{year}{2000}).

\bibitem{mo12}
\bibinfo{author}{Mo, S.-K.} \emph{et~al.}
\newblock \bibinfo{title}{Emerging coherence with unified energy, temperature,
  and lifetime scale in heavy fermion {Y}b{R}h$_2${S}i$_2$}.
\newblock \emph{\bibinfo{journal}{Phys.\ Rev.\ B}}
  \textbf{\bibinfo{volume}{85}}, \bibinfo{pages}{241103(R)}
  (\bibinfo{year}{2012}).

\bibitem{kum15}
\bibinfo{author}{Kummer, K.} \emph{et~al.}
\newblock \bibinfo{title}{Temperature-independent {F}ermi surface in the
  {K}ondo lattice {Y}b{R}h$_2${S}i$_2$}.
\newblock \emph{\bibinfo{journal}{Phys.\ Rev.\ X}}
  \textbf{\bibinfo{volume}{5}}, \bibinfo{pages}{011028} (\bibinfo{year}{2015}).

\bibitem{zwi11}
\bibinfo{author}{Zwicknagl, G.}
\newblock \bibinfo{title}{Field-induced suppression of the heavy-fermion state
  in {Y}b{R}h$_2${S}i$_2$}.
\newblock \emph{\bibinfo{journal}{J. Phys.\ Condens.\ Matter}}
  \textbf{\bibinfo{volume}{23}}, \bibinfo{pages}{094215}
  (\bibinfo{year}{2011}).

\bibitem{hart10}
\bibinfo{author}{Hartmann, S.} \emph{et~al.}
\newblock \bibinfo{title}{Thermopower evidence for an abrupt {F}ermi surface
  change at the quantum critical point of {Y}b{R}h$_2${S}i$_2$}.
\newblock \emph{\bibinfo{journal}{Phys.\ Rev.\ Lett.}}
  \textbf{\bibinfo{volume}{104}}, \bibinfo{pages}{096401}
  (\bibinfo{year}{2010}).

\bibitem{kim06}
\bibinfo{author}{Kimura, S.} \emph{et~al.}
\newblock \bibinfo{title}{Optical observation of non-{F}ermi-liquid behavior in
  the heavy fermion state of {Y}b{R}h$_2${S}i$_2$}.
\newblock \emph{\bibinfo{journal}{Phys.\ Rev.\ B}}
  \textbf{\bibinfo{volume}{74}}, \bibinfo{pages}{132408}
  (\bibinfo{year}{2006}).

\bibitem{col85}
\bibinfo{author}{Coleman, P.}, \bibinfo{author}{Anderson, P.~W.} \&
  \bibinfo{author}{Ramakrishnan, T.~V.}
\newblock \bibinfo{title}{Theory for the anomalous {H}all constant of
  mixed-valence systems}.
\newblock \emph{\bibinfo{journal}{Phys.\ Rev.\ Lett.}}
  \textbf{\bibinfo{volume}{55}}, \bibinfo{pages}{414--417}
  (\bibinfo{year}{1985}).

\bibitem{ong91}
\bibinfo{author}{Chien, T.~R.}, \bibinfo{author}{Wang, Z.~Z.} \&
  \bibinfo{author}{Ong, N.~P.}
\newblock \bibinfo{title}{Effect of {Z}n impurities on the normal-state {H}all
  angle in single-crystal {YB}a$_2${C}u$_{3-x}${Zn}$_x${O}$_{7 - \delta}$}.
\newblock \emph{\bibinfo{journal}{Phys.\ Rev.\ Lett.}}
  \textbf{\bibinfo{volume}{67}}, \bibinfo{pages}{2088--2091}
  (\bibinfo{year}{1991}).

\bibitem{pfau12}
\bibinfo{author}{Pfau, H.} \emph{et~al.}
\newblock \bibinfo{title}{Thermal and electrical transport across a magnetic
  quantum critical point}.
\newblock \emph{\bibinfo{journal}{Nature}} \textbf{\bibinfo{volume}{484}},
  \bibinfo{pages}{493--497} (\bibinfo{year}{2012}).

\bibitem{oes08}
\bibinfo{author}{Oeschler, N.} \emph{et~al.}
\newblock \bibinfo{title}{Low-temperature specific heat of
  {Y}b{R}h$_2${S}i$_2$}.
\newblock \emph{\bibinfo{journal}{Physica B}} \textbf{\bibinfo{volume}{403}},
  \bibinfo{pages}{1254--1256} (\bibinfo{year}{2008}).

\bibitem{sen08}
\bibinfo{author}{Senthil, T.}
\newblock \bibinfo{title}{Critical {F}ermi surfaces and non-{F}ermi liquid
  metals}.
\newblock \emph{\bibinfo{journal}{Phys.\ Rev.\ B}}
  \textbf{\bibinfo{volume}{78}}, \bibinfo{pages}{035103}
  (\bibinfo{year}{2008}).

\bibitem{sun13}
\bibinfo{author}{Sun, P.} \& \bibinfo{author}{Steglich, F.}
\newblock \bibinfo{title}{Nernst effect: Evidence of local {K}ondo scattering
  in heavy fermions}.
\newblock \emph{\bibinfo{journal}{Phys.\ Rev.\ Lett.}}
  \textbf{\bibinfo{volume}{110}}, \bibinfo{pages}{216408}
  (\bibinfo{year}{2013}).

\bibitem{pie05}
\bibinfo{author}{Pietri, R.}, \bibinfo{author}{Rotundu, C.~R.},
  \bibinfo{author}{Andraka, B.}, \bibinfo{author}{Daniels, B.~C.} \&
  \bibinfo{author}{Ingersent, K.}
\newblock \bibinfo{title}{Absence of {K}ondo lattice coherence effects in
  {C}e$_{0.6}${L}a$_{0.4}${P}b$_3$: {A} magnetic-field study}.
\newblock \emph{\bibinfo{journal}{J.\ Appl.\ Phys.}}
  \textbf{\bibinfo{volume}{97}}, \bibinfo{pages}{10A510}
  (\bibinfo{year}{2005}).

\bibitem{lee06}
\bibinfo{author}{Lee, P.~A.}, \bibinfo{author}{Nagaosa, N.} \&
  \bibinfo{author}{Wen, X.-G.}
\newblock \bibinfo{title}{Doping a {M}ott insulator: Physics of
  high-temperature superconductivity}.
\newblock \emph{\bibinfo{journal}{Rev.\ Mod.\ Phys.}}
  \textbf{\bibinfo{volume}{78}}, \bibinfo{pages}{17--85}
  (\bibinfo{year}{2006}).

\bibitem{ram15}
\bibinfo{author}{Ramshaw, B.~J.} \emph{et~al.}
\newblock \bibinfo{title}{Quasiparticle mass enhancement approaching optimal
  doping in a high-${T}_c$ superconductor}.
\newblock \emph{\bibinfo{journal}{Science}} \textbf{\bibinfo{volume}{348}},
  \bibinfo{pages}{317--320} (\bibinfo{year}{2015}).

\bibitem{oike15}
\bibinfo{author}{Oike, H.}, \bibinfo{author}{Miyagawa, K.},
  \bibinfo{author}{Taniguchi, H.} \& \bibinfo{author}{Kanoda, K.}
\newblock \bibinfo{title}{Pressure-induced {M}ott transition in an organic
  superconductor with a finite doping level}.
\newblock \emph{\bibinfo{journal}{Phys.\ Rev.\ Lett.}}
  \textbf{\bibinfo{volume}{114}}, \bibinfo{pages}{067002}
  (\bibinfo{year}{2015}).
\end{thebibliography}
\end{document}